\documentclass[sigconf,10pt,screen,authorversion]{acmart}

\usepackage[utf8]{inputenc}
\usepackage[T1]{fontenc}
\usepackage[english]{babel}

\microtypecontext{spacing=nonfrench}

\usepackage{glossaries}
\glsdisablehyper
\glsdisablehyper

\newacronym{acpi}{ACPI}{Advanced Configuration and Power Interface}
\newacronym{acs}{ACS}{ATA Command Set}
\newacronym{afu}{AFU}{Accelerator Function Unit}
\newacronym{ahci}{AHCI}{Advanced Host Controller Interface}
\newacronym{ascii}{ASCII}{American Standard Code for Information Interchange}
\newacronym{asic}{ASIC}{Application-Specific Integrated Circuit}
\newacronym{ata}{ATA}{AT Attachment}
\newacronym{atapi}{ATAPI}{AT Attachment Packet Interface}
\newacronym{atf}{ATF}{ARM Trusted Firmware}
\newacronym{bar}{BAR}{Base Address Register}
\newacronym{bdk}{BDK}{Board Development Kit}
\newacronym{bist}{BIST}{Built-in Self-test}
\newacronym{bmc}{BMC}{Baseboard Management Controller}
\newacronym{bram}{BRAM}{Block RAM}
\newacronym{caam}{CAAM}{Cryptographic Acceleration and Assurance Module}
\newacronym{cam}{CAM}{Common Access Method}
\newacronym{capi}{CAPI}{Coherent Accelerator Processor Interface}
\newacronym{capp}{CAPP}{Coherent Accelerator Processor Proxy}
\newacronym{cca}{CCA}{Confidential Compute Architecture}
\newacronym{ccip}{CCI-P}{Core Cache Interface}
\newacronym{ccix}{CCIX}{Cache Coherent Interconnect for Accelerators}
\newacronym{ccpi}{CCPI}{Cavium Coherent Processor Interconnect}
\newacronym{ci}{CI}{Continuous Intergration}
\newacronym{clut}{CLUT}{Cache Line Under Test}
\newacronym{cpld}{CPLD}{Complex Programmable Logic Device}
\newacronym{cpu}{CPU}{Central Processing Unit}
\newacronym{cxl}{CXL}{Compute eXpress Link}
\newacronym{dac}{DAC}{Digital Analog Converter}
\newacronym{dc}{DC}{Display Controller}
\newacronym{dep}{DEP}{Data Execution Prevention}
\newacronym{dma}{DMA}{Direct Memory Access}
\newacronym{dos}{DoS}{denial-of-service}
\newacronym{dsp}{DSP}{Digital Signal Processor}
\newacronym{eci}{ECI}{Enzian Coherence Interface}
\newacronym{edma}{eDMA}{Enhanced Direct Memory Access}
\newacronym{ept}{EPT}{Extended Page Table}
\newacronym{esai}{ESAI}{Enhanced Synchronous Audio Interface}
\newacronym{esp}{ESP}{Executable Space Protection}
\newacronym{fat}{FAT}{File Allocation Table}
\newacronym{fis}{FIS}{Frame Information Structure}
\newacronym{fiu}{FUI}{FPGA Interface Unit}
\newacronym{fmc}{FMC}{FPGA Mezzanine Card}
\newacronym{fpga}{FPGA}{Field-Programmable Gate Array}
\newacronym{fsis}{FSIS}{Filesystem Information Sector}
\newacronym{fsm}{FSM}{Finite State Machine}
\newacronym{gpu}{GPU}{Graphics Processing Unit}
\newacronym{gqr}{GQR}{Generalized-Quasi-Reduction}
\newacronym{gsync}{\textsc{GSync}}{Global Synchronization}
\newacronym{hba}{HBA}{Host Bus Adapter}
\newacronym{hbm}{HBM}{High-Bandwidth Memory}
\newacronym{hpc}{HPC}{High-Performance Computing}
\newacronym{i2c}{I\textsuperscript{2}C}{Inter-Integrated Circuit}
\newacronym{ich}{ICH}{I/O Controller Hub}
\newacronym{ic}{IC}{Integrated Circuit}
\newacronym{idc}{IDC}{Inter-Domain Communication}
\newacronym{iee}{IEE}{Inline Encryption Engine}
\newacronym{ipi}{IPI}{Inter-Processor Interrupt}
\newacronym{ip}{IP}{intellectual property block}
\newacronym{ipmi}{IPMI}{Intelligent Platform Management Interface}
\newacronym{iut}{IUT}{Implementation Under Test}
\newacronym{kaslr}{KASLR}{Kernel Address Space Layout Randomization}
\newacronym{lba}{LBA}{Logical Block Address}
\newacronym{llc}{LLC}{Last-Level Cache}
\newacronym{lru}{LRU}{Least-recently used}
\newacronym{mmio}{MMIO}{Memory-Mapped I/O}
\newacronym{mmu}{MMU}{Memory Management Unit}
\newacronym{mpsoc}{MPSoC}{Multiprocessor System-on-a-Chip}
\newacronym{mpu}{MPU}{Memory Protection Unit}
\newacronym{mpx}{MPX}{Memory Protection Extensions}
\newacronym{msi}{MSI}{Message-Signalled Interrupt}
\newacronym{mu}{MU}{Messaging Unit}
\newacronym{ndp}{NDP}{Near-Data Processing}
\newacronym{ncq}{NCQ}{Native Command Queueing}
\newacronym{nic}{NIC}{Network Interface Adaptor}
\newacronym{nmp}{NMP}{Near-Memory Processing}
\newacronym{numa}{NUMA}{Non-Uniform Memory Access}
\newacronym{nvme}{NVMe}{NVM Express}
\newacronym{ocapi}{OpenCAPI}{Open Coherent Accelerator Processor Interface}
\newacronym{oci}{OCI}{Octeon III multi-node Interconnect}
\newacronym{os}{OS}{Operating System}
\newacronym{pae}{PAE}{Physical Address Extensions}
\newacronym{pata}{PATA}{Parallel ATA}
\newacronym{pcb}{PCB}{Printed Circuit Board}
\newacronym{pcie}{PCIe}{PCI Express}
\newacronym{pci}{PCI}{Peripheral Component Interconnect}
\newacronym{piix}{PIIX}{PCI IDE ISA Xcelerator}
\newacronym{pim}{PIM}{Processing In Memory}
\newacronym{pio}{PIO}{Programmed I/O}
\newacronym{pmbus}{PMBus}{Power Management Bus}
\newacronym{prd}{PRD}{Physical Region Descriptor}
\newacronym{psci}{PSCI}{Power State Coordination Interface}
\newacronym{psl}{PSL}{POWER Service Layer}
\newacronym[longplural={Page Table Entries}]{pte}{PTE}{Page Table Entry}
\newacronym{qpi}{QPI}{QuickPath Interconnect}
\newacronym{rdma}{RDMA}{Remote Direct Memory Access}
\newacronym{rfis}{RFIS}{Received FIS}
\newacronym{rpc}{RPC}{Remote Procedure Call}
\newacronym{rtc}{RTC}{Real Time Clock}
\newacronym{sai}{SAI}{Synchronous Audio Interface}
\newacronym{sata}{SATA}{Serial ATA}
\newacronym{sbc}{SBC}{single board computer}
\newacronym{scsi}{SCSI}{Small Computer System Interface}
\newacronym{scu}{SCU}{System Controller Unit}
\newacronym{seco}{SECO}{Security Controller}
\newacronym{sgx}{SGX}{Software Guard Extensions}
\newacronym{simd}{SIMD}{Single Input Multiple Data}
\newacronym{sim}{SIM}{SCSI Interface Module}
\newacronym{skb}{SKB}{System Knowledgebase}
\newacronym{smbus}{SMBus}{System Management Bus}
\newacronym{smc}{SMC}{Secure Monitor Call}
\newacronym{smm}{SMM}{System Management Mode}
\newacronym{smmu}{SMMU}{System Memory Management Unit}
\newacronym[longplural={Systems-on-Chip}]{soc}{SoC}{System-on-Chip}
\newacronym[longplural={Systems-on-Module}]{som}{SoM}{System-on-Module}
\newacronym{spl}{SPL}{System Protocol Layer}
\newacronym{tap}{TAP}{Test Access Port}
\newacronym{tcb}{TCB}{Trusted Computing Base}
\newacronym{tcm}{TCM}{Tightly Coupled Memory}
\newacronym{tcp}{TCP}{Transmission Control Protocol}
\newacronym{tdp}{TDP}{Thermal Design Power}
\newacronym{tee}{TEE}{Trusted Execution Environment}
\newacronym{tfa}{TF-A}{Trusted Firmware-A}
\newacronym{tlb}{TLB}{Translation Lookaside Buffer}
\newacronym{tpu}{TPU}{Tensor Processing Unit}
\newacronym{ttbr}{TTBR}{Translation Table Base Register}
\newacronym{uart}{UART}{universal asynchronous receiver-transmitter}
\newacronym{uefi}{UEFI}{Unified Extensible Firmware Interface}
\newacronym{upi}{UPI}{Universal Path Interconnect}
\newacronym{usb}{USB}{Universal Serial BUS}
\newacronym{vfpga}{vFPGA}{Virtual FPGA}
\newacronym{vfs}{VFS}{Virtual Filesystem}
\newacronym{vliw}{VLIW}{Very Long Instruction Word}
\newacronym{vpu}{VPU}{Video Processing Unit}
\newacronym{xmpu}{XMPU}{Xilinx Memory Protection Unit}
\newacronym{xppu}{XPPU}{Xilinx Peripheral Protection Unit}
\newacronym{xrdc}{XRDC}{Extended Resource Domain Controller}
\newacronym{thp}{THP}{Transparent Huge Page}
\newacronym{ddio}{DDIO}{Direct Data I/O}
\newacronym{dbms}{DBMS}{Database Management System}
\newacronym{gc}{GC}{Garbage Collection}

\glsunset{os}
\glsunset{fpga}
\glsunset{pci}
\glsunset{pcie}
\glsunset{usb}

\newcommand{\ndp}{\gls{ndp}\xspace}
\newacronym{mcc}{MCC}{memory channel controller}
\newacronym{cp}{CP}{channel program}

\RequirePackage{listings}

\usepackage{xspace}

\newcommand{\MMNDP}{M\textsuperscript{2}NDP\xspace}

\newcommand{\etal}{\textit{et al.}\xspace}

\usepackage{xcolor}
\usepackage{todonotes}

\newcommand*\circled[1]{\raisebox{.5pt}{\textcircled{\raisebox{-.9pt} {#1}}}}

\copyrightyear{2025}
\acmYear{2025}
\setcopyright{cc}
\setcctype{by}
\acmConference[APSys '25]{16th ACM SIGOPS Asia-Pacific Workshop on
Systems}{October 12--13, 2025}{Seoul, Republic of Korea}
\acmBooktitle{16th ACM SIGOPS Asia-Pacific Workshop on Systems (APSys '25),
October 12--13, 2025, Seoul, Republic of
Korea}\acmDOI{10.1145/3725783.3764403}
\acmISBN{979-8-4007-1572-3/2025/10}

\begin{document}

\title{Mainframe-Style Channel Controllers for Modern Disaggregated Memory Systems}

\author{Zikai Liu}
\email{zikai.liu@inf.ethz.ch}
\affiliation{%
    \institution{ETH Zurich}
    \city{Zurich}
    \country{Switzerland}
}
\orcid{0009-0000-5411-9785}

\author{Jasmin Schult}
\email{jasmin.schult@inf.ethz.ch}
\affiliation{%
    \institution{ETH Zurich}
    \city{Zurich}
    \country{Switzerland}
}
\orcid{0009-0000-1815-3206}

\author{Pengcheng Xu}
\email{pengcheng.xu@inf.ethz.ch}
\affiliation{%
    \institution{ETH Zurich}
    \city{Zurich}
    \country{Switzerland}
}
\orcid{0000-0002-2724-7893}

\author{Timothy Roscoe}
\email{troscoe@inf.ethz.ch}
\affiliation{%
    \institution{ETH Zurich}
    \city{Zurich}
    \country{Switzerland}
}
\orcid{0000-0002-8298-1126}

\begin{abstract}

Despite the promise of alleviating the main memory bottleneck, and the
existence of commercial hardware implementations, techniques for
\textit{\acrlong{ndp}} have seen relatively little real-world
deployment.  The idea has received renewed interest with the
appearance of disaggregated or ``far'' memory, for example in the use
of CXL memory pools.

However, we argue that the lack of a clear OS-centric abstraction of
\acrlong{ndp} is a major barrier to adoption of the technology.
Inspired by the \emph{channel controllers} which interface the CPU to
disk drives in mainframe systems, we propose \textit{\acrlongpl{mcc}}
as a convenient, portable, and virtualizable abstraction of
\acrlong{ndp} for modern disaggregated memory systems.

In addition to providing a clean abstraction that enables OS
integration while requiring no changes to CPU architecture,
\acrlongpl{mcc} incorporate another key innovation: they exploit the
cache coherence provided by emerging interconnects to provide a much
richer programming model, with more fine-grained interaction, than has
been possible with existing designs.

\end{abstract}

\begin{CCSXML}
<ccs2012>
   <concept>
       <concept_id>10011007.10010940.10010941.10010949</concept_id>
       <concept_desc>Software and its engineering~Operating systems</concept_desc>
       <concept_significance>500</concept_significance>
       </concept>
   <concept>
       <concept_id>10010520.10010575.10010580</concept_id>
       <concept_desc>Computer systems organization~Processors and memory architectures</concept_desc>
       <concept_significance>500</concept_significance>
       </concept>
   <concept>
       <concept_id>10010520.10010521.10010537</concept_id>
       <concept_desc>Computer systems organization~Distributed architectures</concept_desc>
       <concept_significance>300</concept_significance>
       </concept>
   <concept>
       <concept_id>10010583.10010600.10010628.10010629</concept_id>
       <concept_desc>Hardware~Hardware accelerators</concept_desc>
       <concept_significance>300</concept_significance>
       </concept>
 </ccs2012>
\end{CCSXML}

\keywords{Near-data processing, operating systems, cache coherence, far memory, disaggregation, offloading, accelerators.}

\ccsdesc[500]{Software and its engineering~Operating systems}
\ccsdesc[500]{Computer systems organization~Processors and memory architectures}
\ccsdesc[300]{Computer systems organization~Distributed architectures}
\ccsdesc[300]{Hardware~Hardware accelerators}

\maketitle

\section{Introduction} \label{sec:intro}

Limited bandwidth to main memory, and the high latency of main memory
accesses relative to CPU frequency, have long been performance
bottlenecks despite the widespread use of techniques to hide or
reduce memory access times (caches, prefetchers, multi-threaded cores,
out-of-order execution, etc.).  This in turn has led to a long line of
architecture research on \gls{pim}, \gls{nmp}, or \gls{ndp}, whereby
processing elements are placed close to main memory and act as offload
engines for some CPU tasks.  However, to date, very little of this work has
found its way to product, and almost nothing to widespread deployment.

Recently, the deployment of \emph{disaggregated}, \emph{pooled}, or
\emph{far memory}, often via new interconnect protocols like
\gls{cxl}~\cite{CXL31}, has given new impetus to this idea, since far memory
incurs significantly higher access latency and delivers lower throughput than
local DRAM~\cite{Li:Pond:2023,Liu:CXLPerf:2025,Sun:2023:DemystifyingCXLMemory}.
For instance, Marvell recently introduced Structera A, a series of
near-memory accelerators with Arm Neoverse V2 cores positioned next to
CXL-attached memory~\cite{Marvell:StructeraA:2024}. 

We contend that the failure of \ndp (which we will henceforth use as
an umbrella term encompassing \gls{pim}, \gls{nmp}, and related
techniques) to impact practice is due to the lack of an OS-centric
perspective on the technique~\cite{Barbalace:NDP-OS:2017}.  Such a
perspective on \ndp would balance, on the one hand, hardware
constraints and improvements in application efficiency with, on the
other, the oft-neglected requirements for complete realistic systems: 
secure multiplexing, resource management, virtualization, portable
abstractions, and so on.

In this paper we develop such an OS-centric view and its implications,
focusing on three facets.  The first is the
\textbf{application model}: the abstractions presented
to application writers.  We argue these abstractions should not only
be easily usable (by both application programmers and compiler
writers) and efficient, but also portable, fully virtualized, and free
from arbitrary resource limitations.

Second is the \textbf{system software model}, in other words the
non-functional system-wide properties we care about: \ndp should
preserve the application's existing security and isolation properties,
applications using it should not experience livelock or starvation,
etc.

Finally, the \textbf{hardware design} should follow the requirements of
the application and system software models.  To date, most proposals
for \ndp have been largely bottom up, with the broader issues left
unresolved.  This has left it unclear which hardware designs do or do
not make sense in a real computer system with multiple tenants.
Moreover, given the nature of the hardware market, it is desirable to
avoid intrusive changes to CPU architecture: the latter break backward
compatibility and limit deployment.  As we argue in this paper, such
architectural changes are unnecessary.

Based on this perspective, we take inspiration from an old idea:
\emph{channel controllers}.  In additional to the main processors, IBM
mainframes have long had processors dedicated to I/O
operations~\cite{Layer:Sys3X0:2003,IBM:Sys360PoO:1964,Padegs:ChannelDesign:1964}.
These processors are typically at least as powerful as the CPU itself
and come with a clean programming abstraction, the \emph{channel
program}~\cite{Padegs:ChannelDesign:1964,IBM:IBM2314:1969}. Applications
like the DB2 relational database include highly-tailored channel
programs as part of the main executable. Under the coordination of the
OS, mainframe channel controllers can serve multiple applications
simultaneously and dramatically cut the overhead of accessing disk
storage.

We revisit this idea but in a very different context: disaggregated
memory systems rather than disks, and the use of cache-coherent access
to far memory (via interconnects such as \gls{cxl}, or
CCIX~\cite{CCIX1}) for both data transfer \emph{and} a control
interface to the remote device.  This permits fine-grained,
low-latency synchronization between an application and an \ndp
accelerator~\cite{Ham:GeneralNDP:2024}.
We present both a clean abstraction of such a \textit{\gls{mcc}}, and a
combined hardware/software design that can efficiently provide such an
abstraction in the context of a modern kernel-based OS.

\section{Background: the memory bottleneck} \label{sec:memory-bottleneck}

Remotely attached, pooled DRAM attached via new interconnects like CXL
are an attractive proposition for data centers, providing elastic
scaling, extended capacity, and even reuse of older DRAM silicon.

However, such flexibility and cost saving comes with a price:
disaggregated memory exhibits higher latency and lower bandwidth than
local DRAM.  Evaluations on existing CXL memory devices 
suggest latency from 150 to 400ns and bandwidth from 18 to
52GB/s~\cite{Li:Pond:2023,Liu:CXLPerf:2025,Sun:2023:DemystifyingCXLMemory}.
Moreover, performance \emph{variance} is also higher compared to local
DRAM. CXL switches additionally introduce a per-hop latency of 200 to
400ns~\cite{Liu:CXLPerf:2025}.  This leads to performance
characteristics which are very different from a classical ``balanced
system''~\cite{Denning:BalancedSys:1969}, and can significantly impact
a range of applications~\cite{Liu:CXLPerf:2025,Li:Pond:2023}.

At the same time, there are compelling reasons to adopt disaggregated
memory for applications which require rapid random access to large amounts of
data, such as in-memory databases.  Such databases can store more data on a
large disaggregated memory pool and share it with multiple compute nodes for
scalability.  However, query execution is typically bottlenecked on
data transfer between the far memory and the CPU
cores~\cite{Zhang:DBMSonDisagg:2020,Liu:CXLPerf:2025}.  
The same argument applies to graph processing workloads: disaggregated memory allows
storing and sharing larger graphs, but many graph algorithms are
highly sensitive to memory
latency~\cite{Lumsdaine:Challenges:2007,Zhang:Polymer:2015} -- indeed,
even local memory is typically a bottleneck for these algorithms, and disaggregated
memory amplifies the problem.

This has led to renewed interest in the long-standing research area of
\gls{ndp} (broadly construed).  For example, \gls{pim} systems have
been proposed for database
acceleration~\cite{Baumstark:PIM4DB:2023,Gomez-Luna:PrIM:2022}, graph
processing~\cite{Ahn:Tesseract:2015,zhuo_graphq_2019} and other
workloads.
In those systems, computing units are integrated with DRAM
chips and can operate on individual DRAM rows and latches.  This
minimizes the distance that data must be moved, but imposes strong,
hardware-specific constraints on how computation can be performed.  For
example, in UPMEM (a commercial PIM device), each
DRAM processing unit (DPU) can only access a fixed 64MB DRAM slice.
This requires complex software on the CPU to steer the data flow to
and from DPUs~\cite{Gomez-Luna:PrIM:2022}. 
From an OS-centric perspective, the memory and computing resources are
too closely-coupled to be virtualizable or to provide anything more
than coarse-grained inter-application isolation.  Instead, the
application developer is required to partition the data for a specific
hardware platform. Virtualizing and multiplexing UPMEM has been
suggested but at a granularity of multiple gigabytes~\cite{Teguia:vPIM:2024a}.

Other \gls{nmp} systems are less architecturally restrictive.
Lockerman \etal~\cite{Lockerman:Livia:2020} suggest adding
compute units \emph{throughout} the memory hierarchy. A downside with this approach
is the pervasive hardware changes it entails.  The additional die area
is small, but the end-to-end usability additionally requires hardware
verification (on every changed component), and the issue of secure
multiplexing the new compute resources is left
unaddressed.

We are not the first one to notice those problems.
Gao~\etal~\cite{Gao:NDPAnalytics:2015} and
Ghose~\etal~\cite{Ghose:PIMAdoption:2019} observe the gaps in address
translation, memory protection and isolation
functionality. Barbalace~\etal~\cite{Barbalace:NDP-OS:2017}
additionally discuss the problems in scheduling and the programming
model, and call for better runtime and OS support.  More recently, Ham~\etal propose
M$^{2}$NDP~\cite{Ham:GeneralNDP:2024}, which focuses on \gls{ndp} for
CXL-based disaggregated memory, emphasizing the limited hardware
changes required as an important factor.

The \gls{ndp} landscape, including recent work using CXL-attached
memory~\cite{Ham:GeneralNDP:2024,Sim:CMS:2022,Huangfu:Beacon:2022,Jan:CXLANNS:2023}
shows a wide range of system designs, with divergent and largely
incompatible programming models and constraints on secure
multiplexing.  We build on this work, but by taking an OS-centric
perspective we hope to create more broadly practical systems,
particularly in the context of disaggregated far memory.

\section{Design overview} \label{sec:abs}

\begin{figure}[tbp]  %
    \centering
    \includegraphics[width=\linewidth]{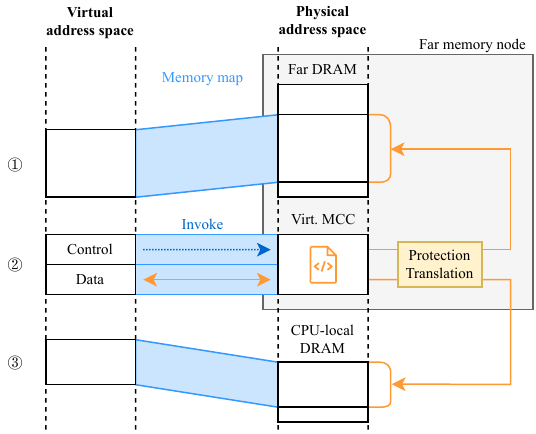}
    \caption{The far memory and \gls{mcc} abstraction}
    \label{fig:memory-map}
\end{figure}

We now discuss our proposals for abstracting \ndp in a portable,
multiplexed, usable manner.
A good interface makes the task of the application
developer easier by providing high-level abstractions, without
compromising performance by allowing the underlying system software
and hardware to work in the most efficient way possible.  It should
also allow a programmer (or code generator) to reason
about the performance impact of using the interface.

A key feature of OS abstractions (like virtual memory, files,
or sockets) is that they impose no arbitrary limits on usage:
modulo complete (and rare) resource exhaustion, a
program is always free to create a new file, extend an existing one,
use more virtual memory, etc.  When acquiring processing resources
close to far memory, this means that a \emph{user program should not
be limited by a fixed number of hardware units}.
This feature can also be viewed as one case of a general notion of
\emph{portability}: code written to use one hardware platform should,
ideally, run correctly on a different platform. 

These requirements lead us to abstract \ndp processes as a set of
virtual, dynamically instantiated processors (\glspl{mcc}) close to
memory, with a standardized interface.  In order to both take
advantage of existing OS abstractions, and to exploit the new
possibilities presented by cache-coherent interconnects for
fine-grained interaction between CPU threads and \ndp resources, we
build the \glspl{mcc} abstraction over a process' virtual address
space, as shown in \autoref{fig:memory-map}. 

First, application code on the CPU might have direct access to a
region of far memory mapped (via the MMU) to a contiguous region of
the virtual address space~(\circled{1}).  Access to far memory in this way
completely bypasses \ndp resources, but nevertheless this illustrates
a key design decision: different remote memory nodes map to
different regions of the virtual address space, and thus \emph{the
physical location of memory is explicit in the virtual address space
layout}.

This is in stark contrast to transparent tiered memory systems like
Intel Flat Memory Mode~(FMM)~\cite{Zhong:2024:ManagingMemoryTiers},
TPP~\cite{Maruf:TPP:2023} and M5~\cite{Sun:M5:2025}, where data
placement is not exposed to the applications, but similar to that of
\MMNDP~\cite{Ham:GeneralNDP:2024} and
\textsc{CtXnL}~\cite{Wang:CTXNL:2025}.

While there may be advantages to hiding memory properties from legacy
applications which simply want to exploit expanded memory, efficient
use of \ndp strongly motivates making the distributed nature of far
memory explicit.  For memory-intensive applications like databases and
graph processing, software has a better knowledge of data placement
and access patterns, while hardware can only speculate.  Past
experience suggests transparent features like FMM end up being
bypassed: for example, DBMSs generally disable \gls{thp}
support~\cite{RedisDiagnosingLatency} and graph applications manage
NUMA memory explicitly~\cite{Zhang:Polymer:2015}.

Second, each \gls{mcc} occupies its own region of virtual
address space~(\circled{2}), and a user program communicates with its
private \glspl{mcc} using memory operations on these
regions, as described in the next section. 
The control and data flow are indicated by the two arrows in the region in
\autoref{fig:memory-map}. 

Finally, each \gls{mcc} has the ability to access both far memory and host-local
memory belonging to its application in a conventional way using \gls{dma}
operations~(\circled{3}), illustrated by the other two orange arrows.

An application wishing to use \ndp with far memory first acquires
access to the relevant region(s) (for example, on Linux via a variant
of \texttt{mmap()}), and then asks the kernel to instantiate one or
more \glspl{mcc}, resulting in the creation of a new region
(\circled{2}) of virtual address space for each one.  Since each
\gls{mcc} is private to an application, and its access rights are
limited by the OS and hardware to the application's virtual address
space, the application can assume protection and operate as if it had
exclusive access. 

\section{The \gls{mcc} abstraction}

The central contribution of this work is the nature of the interface to an
\gls{mcc} over virtual memory accesses.
The \gls{mcc} memory region is divided into two areas, one for
control, and one for data.  The control area allows the application to
configure the \gls{mcc} directly over \gls{mmio} without involving the
local kernel, including downloading a \emph{channel programs} to the
\gls{mcc}.

A \gls{cp} differs significantly from previous models for programming
\ndp accelerators.  In a typical existing approach, the \ndp resource
is given a (potentially lengthy) task to perform (such as zeroing
memory, or materializing a view in an in-memory database).  It runs to
completion, potentially reading and writing both local (to itself)
memory and host memory.

The \gls{cp} programming model, however, also includes ongoing
interactions with the host application, which occur via transactions
using the cache coherence protocol in the data area.  For example, a
\gls{cp} might \emph{stream} results packed in cache lines directly to
the cache running the application, using coherence for synchronization
and message passing as in~\cite{Ruzhanskaia:PIO:2025}, or present an
ongoing query-style interface to a remote data structure using
cache-line-sized reads and writes as in~\cite{Ham:GeneralNDP:2024}.
The \gls{mcc} abstraction therefore exposes, in the form of a
\gls{cp}, a much richer programming model to applications.

One advantage of this approach is performance: it is hard to beat the
coherence protocol for latency in transferring data units up to a few
kilobytes in size, and it also eliminates the overhead of setting up a
new ``task'' for each \ndp operation to be performed by the
application.

In addition, it provides a \emph{logical view} of data produced by the
\gls{mcc}.  The application issues loads and stores to this region,
but instead of accessing physical memory, the \gls{cp} generates responses
programmatically at runtime.

At the same time, it avoids the portability and compatibility issues,
and also the security risks, that accompany extending the processor
architecture to communicate with far memory, as in Intel's new AiA
interface~\cite{Intel:CVE202421823:2024,ArjanvandeVen:VFIO:2024}.

The precise semantics for \glspl{cp} is, at this point, an open
question, but a promising approach is, at the lowest level, an
event-driven model where the events include the arrival of coherence
messages from the host (typically triggered by load and store
operations in application code) and completions of local operations to
access DRAM.  However, the programming model exposed to application
writers should clearly hide most of these details.

Another open question is whether an \gls{mcc} should be restricted to
accessing memory on a single far memory node (typically, the one where
the hardware that it runs on is located), or if should see anything
accessible from the application's virtual address space.  In the
latter case, we argue it is still important that each \gls{mcc} has an
\emph{affinity} which specifies what memory is local to the \gls{mcc}.
When the system scales to multiple memory nodes, data placement strategies
become relevant for reducing data movement across nodes (potentially
with additional latency due to interconnect switches).

\section{System design issues} \label{sec:impl}
  
\begin{figure}[tbp]  %
    \centering
    \includegraphics[width=\linewidth]{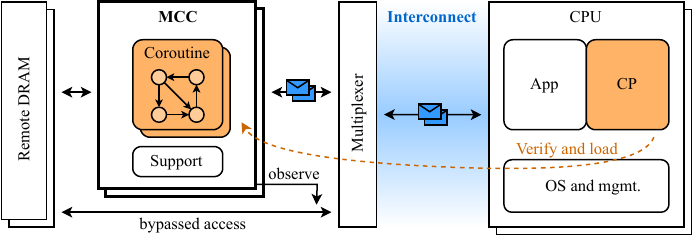}
    \caption{System architecture}
    \label{fig:architecture}
\end{figure}

We now discuss implementation issues in implementing
the \glspl{mcc} abstraction, summarized shown in \autoref{fig:architecture}.

\subsection{Assumptions on the interconnects}

The system is designed around the emerging interconnects with
memory access semantics. While we do not bound the design to a specific
protocol, we make several assumptions.

First, the interconnect is message-based and encodes \textbf{memory
  transactions in a fixed-size granule} (usually a cache line). This is
the case for \gls{cxl}, \gls{ccix}, and the \gls{eci} on the Enzian research
platform~\cite{Cock:Enzian:2022,Ramdas:CCKit:2023} we are using to
prototype the design.
In contrast, \gls{pcie}, originally designed for peripheral devices
with limited intelligence, is oriented heavily towards
device-initiated bulk \gls{dma} transfers and word-sized CPU-initiated
programmed I/O, which is a poor fit for encoding memory transactions.

Furthermore, we focus on protocols that allow \textbf{symmetric coherency},
meaning each party can actively control the cache line ownership symmetrically.
This is \emph{not} the case for the bias-based \gls{cxl}.cache protocol, where a
device needs to query the host CPU to resolve the coherency, incurring 
an extra round-trip delay.   \gls{cxl}.mem 3.0 \emph{does} have this
property by including a back-invalidation channel, allowing the device
to independently resolve coherence with a directory (or ``snoop
filter'' in \gls{cxl} terms).  However, to date no implementations yet exist. 
However, \gls{eci}, originally designed as a coherence interconnect
for CPU sockets, also adopts this model, with
performance~\cite{Cock:Enzian:2022,Ramdas:CCKit:2023} that is
comparable with \gls{cxl}.

Symmetric coherence is important for our design as it allows \gls{mcc} to
perform fine-grained data movement independently. However, we note that the
need for coherence \emph{mechanisms} does not imply \emph{full coherency} on
the whole memory space. With the knowledge of high-level memory access patterns,
\gls{mcc} and \gls{cp} can eliminate unnecessary coherence traffic while
maintaining application correctness.

\subsection{\gls{mcc} execution environment}

The physical \gls{mcc} complex consists of general-purpose processors
(albeit not as powerful as mainframe channel controllers) and a set of
supporting units: a fast scratchpad memory, a split-phrase copy engine
between scratchpad and DRAM, and streaming interfaces to coherence
through which the processor receives and responds to coherence messages.
The scratchpad can be accessed locally in a single
cycle. Between it and the DRAMs, data movement is performed through
the copy engine under the instruction of the processor. The transfer
can be initiated in one cycle and is asynchronous, allowing the \gls{cp} to
hide the data movement latency with proper instruction scheduling. 

A key implication of our OS-centric abstraction is
that it is fully virtual: there will be a fixed number of physical
processors on a node for running \glspl{cp}, but an unbounded number
of \glspl{mcc}, each of which is dedicated to an application.  
This imposes several requirements for the processor and the supporting system
software. 

First, a physical processor on a far memory node must
\textbf{multiplex} a number of \glspl{mcc}, which also implies that it
must perform \textbf{scheduling}.  Given that \glspl{cp} are specified
at a high level, this scheduling need not be preemptive -- effectively
the physical processor can cooperatively schedule \gls{cp}
interpreters as \emph{coroutines} without sacrificing performance.
While relatively simple scheduling might provide sufficient guarantees
against starvation under load, we might consider more complex policies
(such as weighted fair queuing for memory and interconnect access).

In addition to multiplexing, \glspl{mcc} must provide
\textbf{isolation and memory protection}.  Effectively, this means
that the physical processors on the far memory node must be kept
up-to-date with the virtual address spaces of applications on the host
CPU.

At first sight, this would seem to introduce a high system overhead,
but we make two observations.  First, the entire address space need
not be replicated on the channel controller, and restricting far
memory to contiguous mappings further simplifies the metadata that
must be kept consistent, effectively
segmentation~\cite{Bensoussan:MulticsVM:1972,Chiueh:SegAndPaging:1999}.
Second, we are encouraged by recent
proposals for fine-grained synchronization between CPU-based OSes and
network interfaces on a coherent interconnect~\cite{lauberhorn}, which
suggest that even scheduling state can be efficiently shared using
the same techniques~\cite{Ham:GeneralNDP:2024,Ruzhanskaia:PIO:2025} we
propose for data transfer.
Furthermore, extending on the same interface to the interconnect, 
\gls{mcc} can also observe
memory requests from CPUs, which enables another group of applications
(\autoref{sec:workloads}).

We note that, while we are proposing a minimum level of functionality
on the far memory nodes (i.e. processors capable of scheduling multiple
channel programs, some memory protection, and low-level access to the
coherence protocol), we do \emph{not} require any changes to the CPU
architecture or memory interface, nor to the interconnect protocols.
The cost of the additional hardware can be amortized by
the forthcoming far memory controllers that need to be designed and
manufactured anyway.

\subsection{Preliminary \gls{cp} design}

The \gls{cp} model design is ongoing work. We are
experimenting with a design where the \gls{cp} polls coherence messages on
the \gls{mcc}-specific region (\circled{2} in \autoref{fig:memory-map}),
performs computations, and responds with coherence messages if needed.

When a reply message is needed (e.g. memory read transactions), the
\gls{cp} is on the critical path. If it does not produce an output in
time, the interconnect can be deadlocked, which makes a hard real-time
problem. This is challenging, but we believe it is solvable, following
the line of hard real-time systems that have already been built
(e.g. avionic flight controllers). Specific to our design, there are
two additional advantages. First, latencies of operations, such as
fetching a cache line from the DRAM controller, are highly predictable.
Second, interconnects are typically lenient about timeout. Both \gls{cxl}
and \gls{eci} allow timeout up to the millisecond scale.

When considering the \gls{mcc} virtualization, additional handling is
needed. For example, application and \gls{cp} can be \textbf{co-scheduled} to ensure
no in-flight memory transaction can be issued when the \gls{cp} is descheduled. If the
physical \glspl{mcc} are overloaded, using the CPU to simulate the \gls{mcc}
may be a feasible option.

Above this, we are also developing a \textbf{safe high-level programming model},
which needs to hide the hardware execution environment discussed above
and ensure it is used in a safe manner, while
still allowing the application developer to specify the data movement
explicitly. One notable idea is to use a model like
DataPipes~\cite{Vogel:DataPipes:2023}, where data locations are
specified explicitly but the low-level memory operations are
abstracted away. 
The transformation from a high-level \gls{cp} to a correct compiled 
\gls{cp} requires a combination of language design, compiler checks, 
and run-time verification.

\section{Mapping workloads to the system} \label{sec:workloads}

\begin{figure}[tbp]  %
    \centering
    \includegraphics[width=\linewidth]{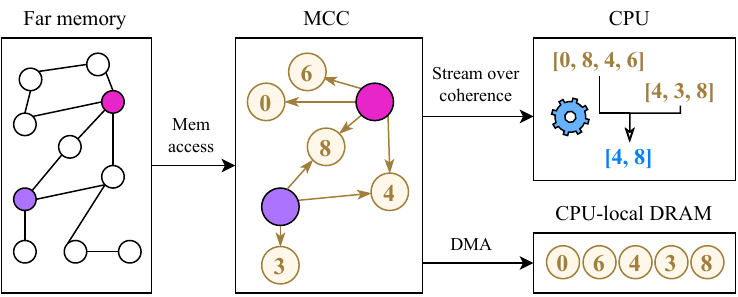}
    \caption{$n$-hop common neighbor pipeline}
    \label{fig:common-neighbors}
\end{figure}

To illustrate how \glspl{mcc} works in practice, consider calculating
common neighbors in a graph (LinkedIn uses this to find
$n$-hop common connections between
users~\cite{Wang:SetCoverLinkedIn:2013}).
\autoref{fig:common-neighbors} shows a way to map the processing
pipeline using an \gls{mcc}: 

\begin{itemize}
\item The complete social graph resides in far memory.
\item The \gls{cp} traverses the graph from each source node specified
  by the CPU and gathers a list of (possibly indirect) neighbors. 
\item This list of node IDs is streamed to CPU caches and then registers over the
  coherence protocol.
\item The second stage (list intersection) is perform on
  the CPU, which now has excellent data locality.
\item In addition, if the CPU needs to access the auxiliary node information, 
  the \gls{cp} can exercise the DMA capability to copy those data to the 
  CPU-local memory.
\end{itemize}

Like many \textbf{graph processing workloads}, this is a good fit for an
\gls{mcc} since it has relatively low computation per node but highly
irregular and unpredictable memory access
patterns~\cite{Lumsdaine:Challenges:2007,Zhang:Polymer:2015}, and so
benefits from the reduced latency from an \gls{mcc} to far memory. 
Other examples include online and query-heavy graph processing
workloads like graph traversal (e.g. BFS), single-source shortest path,
k-nearest neighbors, and pattern matching. These workloads, in their practical
use-cases, tend to have real-time requirements of low latency and high
throughput~\cite{Cheng:TAOBench:2022,Wang:SetCoverLinkedIn:2013}, which
motivates the usage of disaggregated memory and \gls{mcc}.

In contrast, workloads like PageRank~\cite{Page:PageRank:1999} and triangle
counting are less good fits for the system. They require massive computation and
a high degree of parallelism, which is not a key strength of the
\gls{mcc} approach. 
In practice, these workloads benefit more from cluster-based offline
parallel processing, and existing batch processing systems work well.

The example above is one way to use an \gls{mcc} -- splitting a linear
pipeline between \gls{mcc} and CPU.  Another approach is to perform
smart prefetching, an idea well explored by architecture
research~\cite{Yang:DataLinkedList:2000,Solihin:Prefetching:2002}. \gls{mcc},
being a generous-purpose processor, is likely to be somewhat slower
than hardware prefetchers, but much more flexible with regard to data
layouts. Graph applications can bundle \glspl{cp} that understand
the in-memory format of the graph and perform tailored prefetch, while
CPUs perform computations in parallel. 

Another class of workload is \textbf{in-memory databases}. Unlike
graph processing, databases use extensive runtime information on
their own data access patterns and computations. Previous
work~\cite{Korolija:Farview:2021,Woods:Ibex:2014,Jo:YourSQL:2016} has
shown that by pushing query operators closer to data drastically
reduces the memory bottleneck. Korolija~\etal pioneered
offloading query operators to remote memory in
Farview~\cite{Korolija:Farview:2021} using RDMA-based access with an
FPGA as the \ndp unit.  Such an approach would map well to an \gls{mcc}.

Database management systems can include \glspl{cp} or synthesize them at query time to perform
computations near data and/or steer data movement, much as IBM DB2
used mainframe channel controllers.
Moreover, a \glspl{cp} executing part of the operator graph can exploit
different data transfer paradigms to interface with software.  Bulk
transfers of large data sets (e.g. with
low selectivity and/or using eager materialization) benefit from
\gls{dma} to CPU-attached memory, while when operators on a CPU core
can consume intermediate tuples in real time, streaming to the CPU
cache over the coherent interconnect is more efficient.

These exercise the full range of computation and
communication mechanisms available to \glspl{mcc}.  However, there are also
simpler workloads that utilize a smaller feature set. For example, \textbf{bulk
memory copying and clearing} is more efficient on a \gls{mcc} due to its
proximity to far memory, and is important for huge page
zeroing~\cite{Panwar:HawkEye:2019} and copy-on-write~\cite{Kamath:MC2:2024},
hypervisor paging~\cite{Kwon:Ingens:2017,Pham:LargePagesVM:2015}, VM
migration~\cite{Clark:VMLiveMigration:2005,Choudhary:SurveyVMMigration:2017},
and replication for in-memory databases~\cite{MongoDB:InMemoryDB}.

Mapping this workload to \gls{mcc} is intuitive: a simple \gls{cp} is
invoked by the CPU and the \gls{mcc} performs the memory operations
asynchronously. The \glspl{cp} can be parameterized, and the host
application can specify the memory regions at the invocation time
using the control channel.  We expect similar performance benefits to 
previous work~\cite{Kamath:MC2:2024,Yang:Zeroing:2011}.

Finally, an \gls{mcc} can observe memory accesses and assemble
fine-grained \textbf{memory access statistics} for application
software. Such information can be used for hot page migration in tired
memory
systems~\cite{Sun:M5:2025,Duraisamy:TiredMemGoogle:2023,Maruf:TPP:2023},
\gls{gc}~\cite{Chilimbi:GCCache:1998,Courts:GCLocality:1988,Chen:ProactiveGC:2006} 
and profile-guided optimizations~\cite{Chen:AutoFDO:2016,Nagendra:AsmDB:2020}.

\section{Status and conclusion}

An OS perspective on \ndp calls for clean,
portable, and usable abstractions to applications, while providing
inter-application safety, scheduling, and virtualization. It also
helps in mapping the hardware design space that
makes sense in the context of a complete system.

By prototyping our ideas on a real hardware platform, we expect to
establish the programming models, isolation properties, and
application areas that make sense for mitigating the overheads of far
memory.

\bibliographystyle{ACM-Reference-Format}
\bibliography{references}

\end{document}